\documentclass[prl,twocolumn,a4paper]{revtex4}
\usepackage{amssymb}
\usepackage{epsfig}
\usepackage{amsmath}
\usepackage{graphicx}
\usepackage{bm}
\usepackage{times}
\usepackage{color} 
\usepackage{ulem}

\begin{document}

\title{Decoherence assisted transport in a dimer system}

\author{I. Sinayskiy$^{1}$, A. Marais$^{1}$, F. Petruccione$^{1}$ and A. Ekert$^2$}

\affiliation{$^1$ Quantum Research Group, School of Physics and
National Institute for Theoretical Physics, University of KwaZulu-Natal, Durban, 4001, South Africa\\
$^2$ Mathematical Institute, University of Oxford and Centre for
Quantum Technologies, National University of Singapore}

\date{\today}
\begin{abstract}
The dynamics of a dimer coupled to two different environments each in
a spin star configuration under the influence of decoherence
is studied. The exact analytical expression for the transition
probability in the dimer system is obtained for different situations, i.e., independent and correlated environments. In all cases considered, it is shown that there exist well-defined ranges of parameters for which decoherent interaction
with the environment assists energy transfer in the dimer system.
In particular, we find that correlated environments can assist
energy transfer more efficiently than separate baths.
\end{abstract}
\maketitle

The processes of energy and information transfer in quantum
networks play an important role for quantum communication and
quantum computation. In realistic physical situations the unavoidable interaction with the environment leads to decoherence
and dissipation which typically play a coherence-destructive role
\cite{toqs}. However, recently ultrafast spectroscopic techniques have been claimed to reveal long-lasting quantum coherence in biological systems, including the photosynthetic light-harvesting complexes of a species of green sulphur bacteria \cite{Engel,pcn}, a species of purple bacteria \cite{Fl}, and two species of marine cryptophyte algae \cite{Collini}.
Pigment-protein light harvesting antenna in the photosynthetic complex transfer excitonic energy rapidly and efficiently \cite{ritz} through a series of electronic excitations to the reaction centre \cite{Gr}. The efficiency of the energy transfer through the network of chromophores and the evidence of quantum coherence has led to discussion about the role of the environment in the quantum transfer process, and the degree to which it may contribute to the transport efficiency. 


It has been shown that in a simple model of an aggregate of monomers interacting through dipole-dipole forces with realistic coupling strengths, quantum and classical coherent transport are comparable \cite{wave}. At the same time there is much activity proposing mechanisms for environment-assisted excitonic energy transport in quantum networks, including under the broad headings of noise-assisted transport \cite{AG} and oscillation-enhanced transport \cite{osc}. The possibility that quantum entanglement may enhance transport has also been discussed \cite{ent}.

Modeling the complexity of the environment is a challenge \cite{comp}. The protein-solvent environment interacts strongly with the pigments due to its polarity and as a result can have a significant effect on the quantum dynamics \cite{gilm}, which will therefore in general be non-Markovian \cite{ishi}. Such non-Markovian effects have widely been taken into account \cite{ishi, nonM}, but so far, all within spin-boson models of excitons within a protein medium. Any biological system is always in a contact with a bosonic environment. However, the interaction with a more structured environment such as a spin bath is more likely to assist quantum efficiency. Spin baths are natural candidates because the reduced dynamics which they induce are intrinsically non-Markovian \cite{BBP} and the relevance of the electron-nuclear spin interactions in photosynthesis (especially in reaction centers dynamics) has been recognized for a long time \cite{spin_spin}.

In this paper we are going to study the simplest electronic energy
transfer system, namely, a dimer coupled to a spin bath. The Hamiltonian of the dimer is
given by
$H_d=\varepsilon_1|1\rangle\langle1|+\varepsilon_2|2\rangle\langle2|+J\left(|1\rangle\langle2|+|2\rangle\langle1|\right)$,
where $\varepsilon_i$ are the energy levels of the dimer and $J$ is the
amplitude of transition. It is well known that in the absence of an environment, if the initial
excitation is in level $1$, then the maximum of the probability of
transition to level $2$ $\left(P_{1\rightarrow 2}(t)\right)$, will
be given by $\mathrm{Max}\left[P_{1\rightarrow
2}(t)\right]=J^2/(J^2+\Delta^2)$, where $\Delta$ is half of the
energy difference between the levels of a dimer
$\left(\Delta=(\varepsilon_2-\varepsilon_1)/2\right)$. This means
that only in the case $\varepsilon_1=\varepsilon_2$ we can say
that excitation is transferred with certainty
$\left(P_{1\rightarrow 2}(t_0)=1\right)$ at the time moment
$t_0=\pi/2J$. In all other cases
($\varepsilon_1\neq\varepsilon_2$) the probability of transition is
always smaller than one. The aim of this paper is to show that in
a generic case ($\varepsilon_1\neq\varepsilon_2$) for a dimer in
contact with two spin environments, decoherence can enhance energy transfer.

\begin{figure}[t]
\includegraphics[width=0.7\columnwidth]{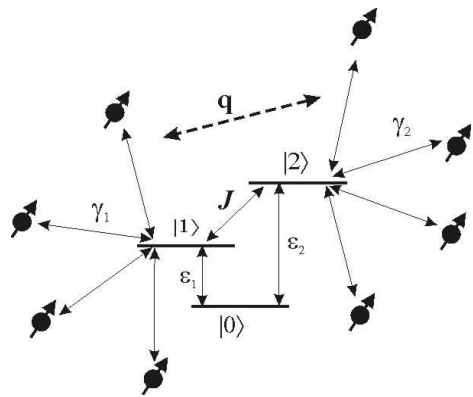}
\caption{Scheme of the dimer coupled to two decoherent
environments in a spin star configuration. A dimer has energy
levels $\varepsilon_1\neq\varepsilon_2$ and amplitude of
transition probability $J$. Each level of the dimer $|1\rangle$
and $|2\rangle$ is in contact with a bath of spins with coupling
constants $\gamma_1$ and $\gamma_2$, respectively. The dashed
double arrow denotes the possibility of correlations between the baths through the Ising-like interaction $\sum_{k,m}\sigma_z^{k,1}\sigma_z^{m,2}$ with strength $q$. }
\end{figure}

For the sake of simplicity and exact solvability the environment
to which the dimer couples will be modeled as a bath of
independent spins $1/2$ in a spin star configuration \cite{BBP}.
Figure 1 depicts the dimer in contact with the two baths. Thus,
the Hamiltonian of the total system has the following form:
\[ H=H_d+H_{B_1}+H_{B_2}+H_{dB_1}+H_{dB_2}.
\]
As it was mentioned above, each environment $B_i$ consists of $N_i$
particles $(i=1,2)$ with spin $1/2$:
\[
H_{B_i}=\alpha_i \sum_{k=1}^{N_i}\frac{\sigma^{k,i}_z}{2},
\]
where $\sigma^{k,i}_z$ are the well-known Pauli matrices. The
decoherent interaction between the dimer and the baths is
described by:
\[H_{dB_i}=\sum_{k=1}^{N_i}\gamma_i |i\rangle\langle
i|\frac{\sigma^{k,i}_z}{2},
\]
where $\gamma_i$ denotes the strength of interaction of the system with
the bath. For a spin bath in such a configuration it is convenient
to define collective spin operators:
\[
S^z_i=\sum_{k=1}^{N_i}\frac{\sigma^{k,i}_z}{2}.
\]
In this notation the total Hamiltonian can be written in the form:
\[
H=J\left(|1\rangle\langle2|+|2\rangle\langle1|\right)+\sum_{i=1}^2\left(\varepsilon_i|i\rangle\langle
i|+\alpha_i S^z_i+\gamma_i|i\rangle\langle i| S^z_i\right).
\]
For the description of the dimer system we express the projectors $|i\rangle\langle j|$ through the Pauli
matrices, i.e.,
\[
|1\rangle\langle 1|=\frac{1_2-\sigma_z}{2}, \quad |2\rangle\langle
2|=\frac{1_2+\sigma_z}{2}, \quad |2\rangle\langle 1|=\sigma^+.
\]
Thus, the total Hamiltonian can be written in the following form:
\[
H=\delta\left(S^z_1,S^z_2\right)1_2+\left(\frac{\varepsilon_2-\varepsilon_1}{2}-\frac{\gamma_1S^z_1-\gamma_2S^z_2}{2}\right)\sigma_z+J\sigma_x.
\]

Let us first consider the simplest special case of both
environments at zero temperature. Obviously, in this case in both
reservoirs all spins are in the ground state. Hence, the initial
state of the bath is a pure state and is described by the
following wave-function:
\[
|\Psi_B(0)\rangle=|\frac{N_1}{2},-\frac{N_1}{2}\rangle\otimes|\frac{N_2}{2},-\frac{N_2}{2}\rangle,
\]
where the vector $|j,m\rangle$ denotes the well known
eigenvectors of the angular momentum operator,
\[
S^2|j,m\rangle=j(j+1)|j,m\rangle,S_z|j,m\rangle=m|j,m\rangle,
\]
$S^2=S^2_x+S^2_y+S^2_z,$ for $j=0,\dots,N/2$ and
$m=-j,\ldots,0,\ldots,j$.

Using the fact that the Hamiltonian of the reservoirs $H_{B_i}$
commutes with the Hamiltonian of the interaction $H_{dB_j}$, the
state of the system will be always of the form:
\[
|\Psi_{\mathrm{Total}}(t)\rangle=\sum_{i=1}^2c_i(t)|i\rangle\otimes|\Psi_B(0)\rangle.
\]
We assume that at time $t=0$ the excitation is in level $1$ of
the dimer, i.e., $c_1(0)=1$ and $c_2(0)=0$. The corresponding
Schr\"odinger equation can be easily integrated and the
probability of transition given by the dynamics of the coefficient
$|c_2(t)|^2$ is found to be
\[
P_{1\rightarrow2}(t)=|c_2(t)|^2=\frac{J^2}{J^2+\Delta^2}\sin^2{\left(t\sqrt{J^2+\Delta^2}\right)},
\]
where \[
\Delta=\frac{\varepsilon_2-\varepsilon_1}{2}+\frac{\gamma_1N_1-\gamma_2N_2}{4}.\]
It is obvious now, that for specially chosen parameters of the
baths ($\gamma_i$ or $N_i$) it is possible to compensate the
energy difference between two levels of the dimer, such that
$\Delta=0$ and $\mathrm{Max}\left[P_{1\rightarrow 2}(t)\right]=1$,
namely by setting
\[
\frac{\varepsilon_1-\varepsilon_2}{2}=\frac{\gamma_1N_1-\gamma_2N_2}{4}.
\]
This implies a very simple mechanism of increasing the probability
of transition in the dimer.

Our main interest, of course, is in the generic case of the dimer
coupled to baths at non-zero temperatures. In this case the
initial state of the bath is given by the canonical distribution
\[
\rho_B(0)=\prod_{i=1}^2\frac{1}{Z_i}e^{-\beta\alpha_iS^z_i},
\]
where $Z_i$ is the partition function of the corresponding bath,
\begin{eqnarray}
Z_i&=&\sum_{j_i=0}^{N_i/2}\sum_{m_i=-j_i}^{j_i}\nu(N_i,j_i)\langle
j_i,m_i|e^{-\beta\alpha_iS^z_i}|j_i,m_i\rangle\nonumber\\ & &
=\sum_{j_i=0}^{N_i/2}\nu(N_i,j_i)\frac{\sinh\beta\alpha_i
\left(j_i+\frac{1}{2}\right)}{\sinh\left(\beta\alpha_i/2\right)},\nonumber
\end{eqnarray}
$\beta$ is the inverse temperature and $\nu(N_i,j_i)$
denotes the degeneracy of the spin bath \cite{molmer, yamen}.
%

%
Using the commutativity of the Hamiltonian of interaction $H_{dB_j}$ and the total Hamiltonian $H$  it is possible to show that in the non-zero temperature case the transition probability is:
\begin{widetext}
\begin{eqnarray}
P_{1\rightarrow2}(t)&=&\mathrm{Tr}_B\left[\langle 2
|U(t)|1\rangle\rho_B(0)\langle 1
|U(t)^\dag|2\rangle\right]\nonumber\\
&=&\frac{1}{Z_1Z_2}\sum_{j_1=0}^{N_1/2}\sum_{m_1=-j_1}^{j_1}\sum_{j_2=0}^{N_2/2}\sum_{m_2=-j_2}^{j_2}\frac{\nu(N_1,j_1)\nu(N_2,j_2)J^2}{J^2+\Delta_{m_1,m_2}^2}\sin^2{\left(t\sqrt{J^2+\Delta_{m_1,m_2}^2}\right)}e^{-\beta\alpha_1m_1-\beta\alpha_2m_2},\nonumber
\end{eqnarray}
\end{widetext}
where $\Delta_{m_1,m_2}$ is given by
\[
\Delta_{m_1,m_2}=\frac{\varepsilon_2-\varepsilon_1}{2}-\frac{\gamma_1m_1-\gamma_2m_2}{2}.
\]

\begin{figure}[t]
\includegraphics[width=0.95\columnwidth]{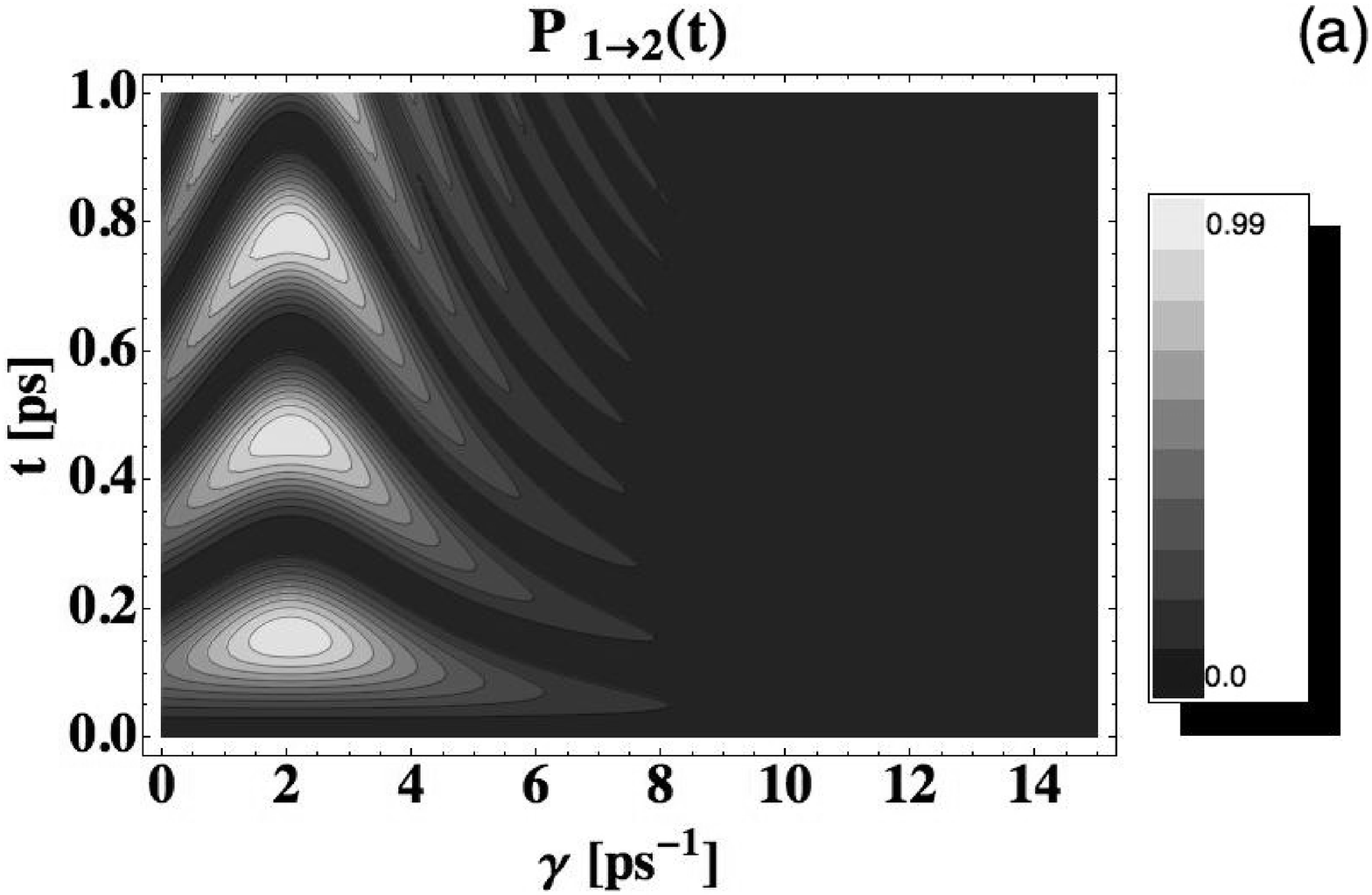}\\
\includegraphics[width=0.95\columnwidth]{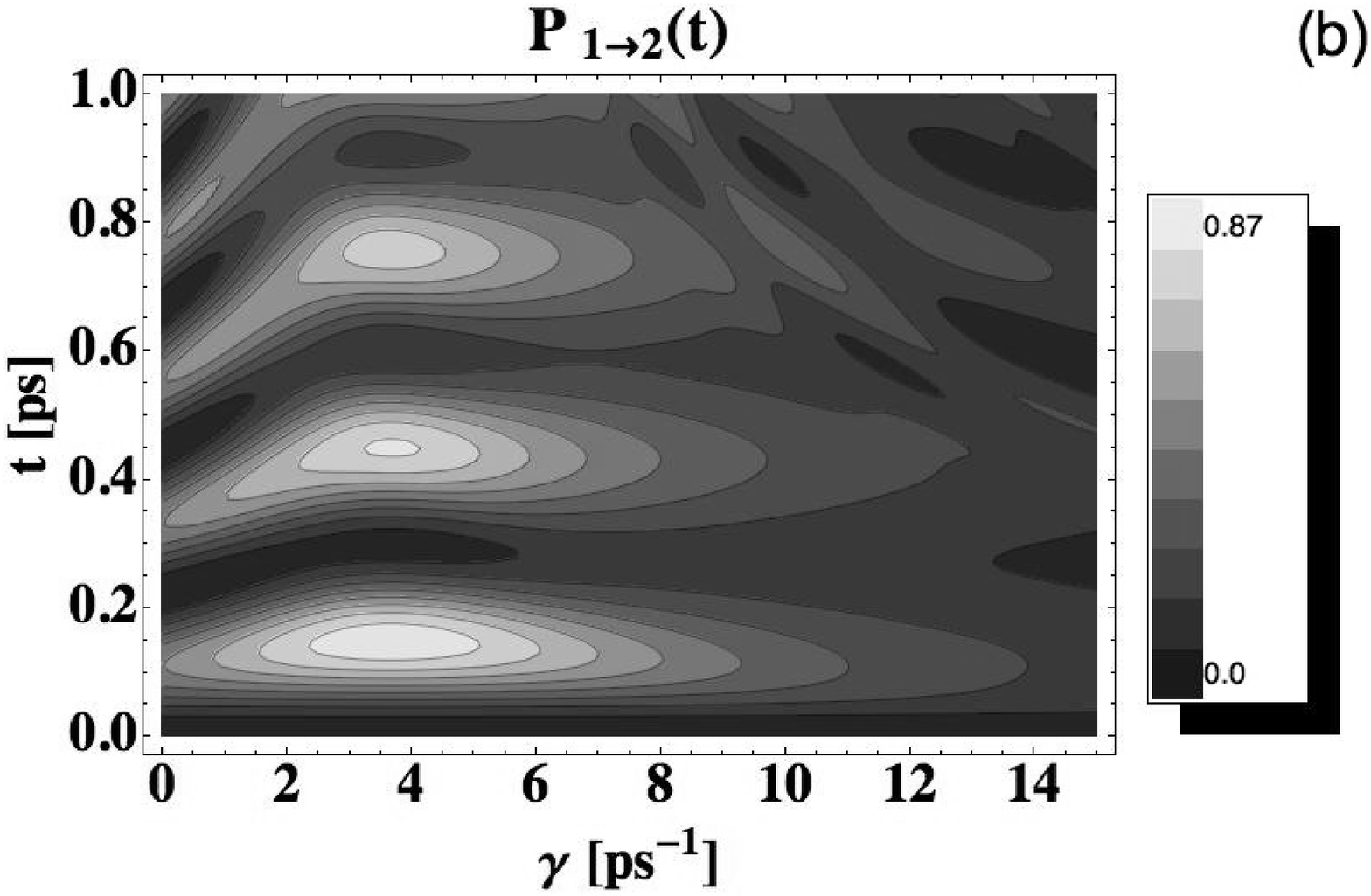}
\caption{Probability of transition $P_{1\rightarrow2}(t)$ in the dimer system with upper level coupled to
a spin bath $\left(\gamma_1=0\right)$ as a function of time
and coupling constant to the spin bath $(\gamma_2=\gamma)$. The temperature of the spin bath in Fig. 2(a) is 77K, while in Fig. 2(b) it is 300K. For both figures the parameters are chosen to
be $N_2=20$, $\alpha_2=250\,\mathrm{ps}^{-1}$, $J=10 \,\mathrm{ps}^{-1}$ and
$\varepsilon_2-\varepsilon_1=20 \,\mathrm{ps}^{-1}$.}
\end{figure}

In Fig. 2 the probability of transition in the dimer system
as a function of time and coupling constant is presented. In this case only the upper level of the dimer is coupled to a spin bath with coupling constant $\gamma_2=\gamma$. In Fig 2(a) we show the results for a spin bath at 77K, whereas in Fig. 2(b) the temperature of the bath is 300K. For both bath temperatures, one can see that the maximum of the
probability of transition is achieved for a non-vanishing interaction with the spin environment. The transfer of energy is enhanced when the bath configurations that contribute to the decrease of the distance between the levels of the dimer prevail.
The increased transition rate occurs on timescales of the order of a few hundred femtoseconds which corresponds to experimental observation \cite{Engel,pcn} and theoretical predictions \cite{photo}. This means that not
only at  zero temperature, but also in more realistic cases,
decoherence assists the energy transfer in the dimer system for the class of models which we are considering here.

The spin bath model allows to investigate analytically the influence of the correlation between environments on the probability of transition. To this
end, we introduce an Ising type interaction between environments with strength $q$, so that the bath Hamiltonian $H_B$ assumes the following form:
\begin{equation}\nonumber
H_B=\alpha_1S^z_1+\alpha_2S^z_2+q S^z_1S^z_2.
\end{equation}

In the special case of environments that are correlated and at
zero-temperature it is easy to see that the probability of
transition is simply given by:
\[
P_{1\rightarrow2}(t)=\frac{J^2}{J^2+\Delta_0^2}\sin^2{\left(t\sqrt{J^2+\Delta_0^2}\right)},
\]
where
\[
\Delta_0=\frac{\varepsilon_2-\varepsilon_1}{2}+\langle \Psi_B(0)|\frac{\gamma_2S^z_2-\gamma_1S^z_1}{2}| \Psi_B(0) \rangle.
\]

The wave-vector $|\Psi_B(0) \rangle$ denotes the initial state of the bath, in this particular case (zero temperature) it would be the ground state of the Hamiltonian $H_B$. Due to the interaction between baths the ground state of the Hamiltonian $H_B$ will depend on parameters $\alpha_1, \alpha_2$ and $q$ of the Hamiltonian $H_B$. We have shown that the state vector $|\Psi_B(0) \rangle$ will be given by
\begin{equation}\nonumber
|\Psi_B(0) \rangle=\begin{cases}
|\frac{N_1}{2},-\frac{N_1}{2}\rangle\otimes|\frac{N_2}{2},-\frac{N_2}{2}\rangle\,\mathrm{for}\,q<q_0,\\
|\frac{N_1}{2},-\frac{N_1}{2}\rangle\otimes|\frac{N_2}{2},\frac{N_2}{2}\rangle\,\mathrm{for}\,q>q_0,\alpha_1>\alpha_2, \\
|\frac{N_1}{2},\frac{N_1}{2}\rangle\otimes|\frac{N_2}{2},-\frac{N_2}{2}\rangle\,\mathrm{for}\,q>q_0,\alpha_1<\alpha_2, \\
\end{cases}
\end{equation}
where $q_0=2\mathrm{Min}\left(\frac{\alpha_1}{N_2},\frac{\alpha_2}{N_1}\right)$. In the case of degeneracy of the parameters, e.g. $q=q_0$ or $\alpha_1=\alpha_2$, one should take the normalized linear combination of the corresponding ground states. For example, if $q>q_0$ and $\alpha_1=\alpha_2$, then the ground state will be given by:
\begin{eqnarray}\nonumber
|\Psi_B(0) \rangle&=&\cos\theta|\frac{N_1}{2},-\frac{N_1}{2}\rangle\otimes|\frac{N_2}{2},\frac{N_2}{2}\rangle\\
& &+\sin\theta e^{i\phi} |\frac{N_1}{2},\frac{N_1}{2}\rangle\otimes|\frac{N_2}{2},-\frac{N_2}{2}\rangle,\nonumber
\end{eqnarray}
 where $0\leq\theta\leq\pi$ and $0\leq\phi<2\pi$.


One can easily formulate conditions under which the interaction with the baths will assist transitions in the system ($\Delta_0=0$), namely
\begin{equation}\nonumber
\frac{\varepsilon_2-\varepsilon_1}{2}={\small\begin{cases}
\left(\gamma_1N_1-\gamma_2N_2\right)/4\, \mathrm{for }\, q<q_0,\\
-\left(\gamma_1N_1+\gamma_2N_2\right)/4\,\mathrm{for}\,q>q_0,\alpha_1>\alpha_2,\varepsilon_2<\varepsilon_1,\\
 \left(\gamma_1N_1+\gamma_2N_2\right)/4\,\mathrm{for}\,q>q_0,\alpha_1<\alpha_2 ,\varepsilon_2>\varepsilon_1.\\
\end{cases}}
\end{equation}
In the most general case considered here, i.e., correlated
environments at non-zero temperature, the probability of
transition is found to be:
\begin{widetext}
\[
P_{1\rightarrow2}(t)=\frac{1}{Z}\sum_{j_1=0}^{N_1/2}\sum_{m_1=-j_1}^{j_1}\sum_{j_2=0}^{N_2/2}\sum_{m_2=-j_2}^{j_2}\frac{\nu(N_1,j_1)\nu(N_2,j_2)J^2}{J^2+\Delta_{m_1,m_2}^2}\sin^2{\left(t\sqrt{J^2+\Delta_{m_1,m_2}^2}\right)}e^{-\beta\alpha_1m_1-\beta\alpha_2m_2-\beta q m_1 m_2},
\]
\end{widetext}
where $\Delta_{m_1,m_2}$ is given by
\[
\Delta_{m_1,m_2}=\frac{\varepsilon_2-\varepsilon_1}{2}+\frac{\gamma_2m_2-\gamma_1m_1}{2}.
\]

\begin{figure}[t]
\includegraphics[width=0.95\columnwidth]{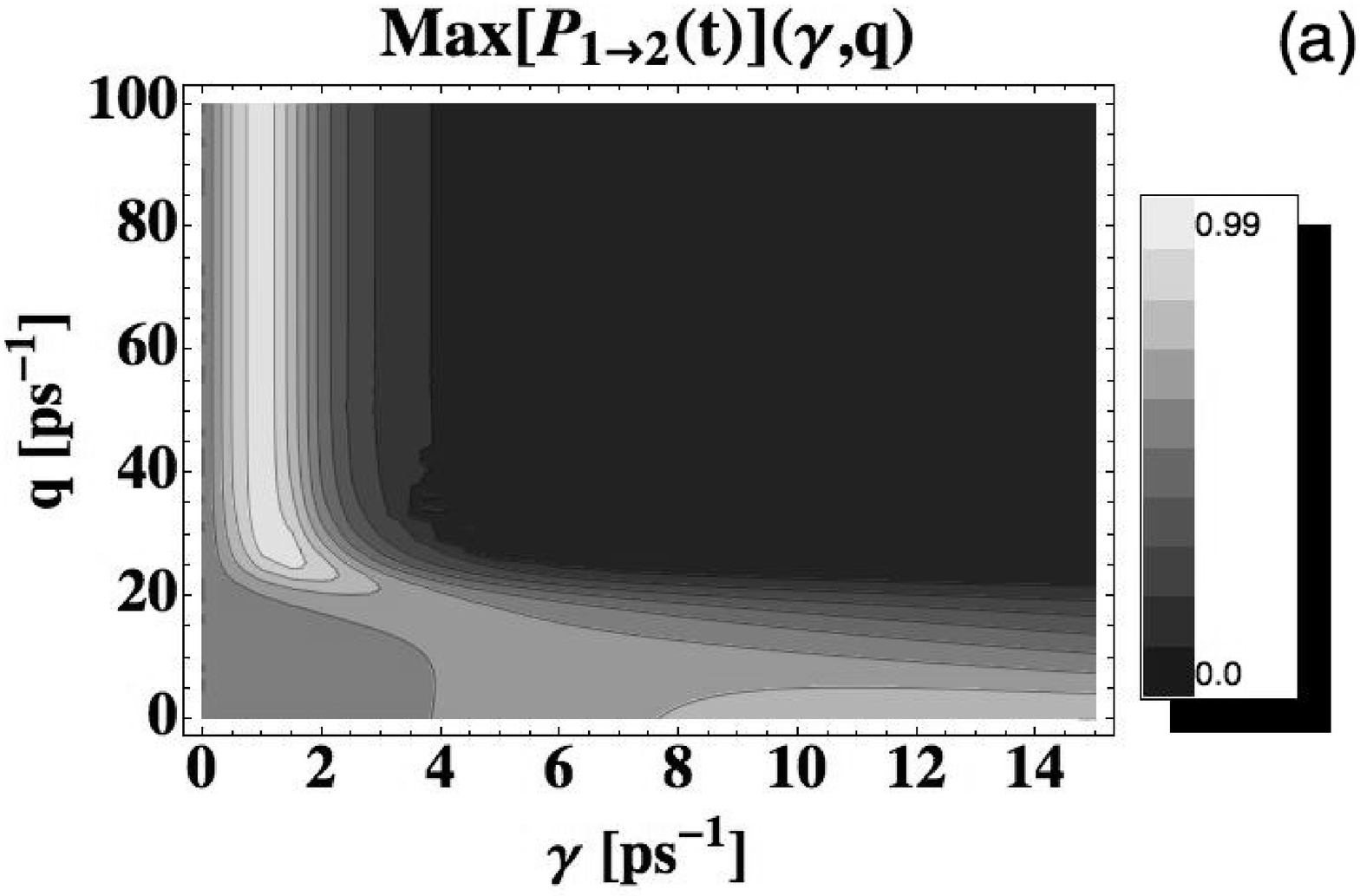}\\
\includegraphics[width=0.95\columnwidth]{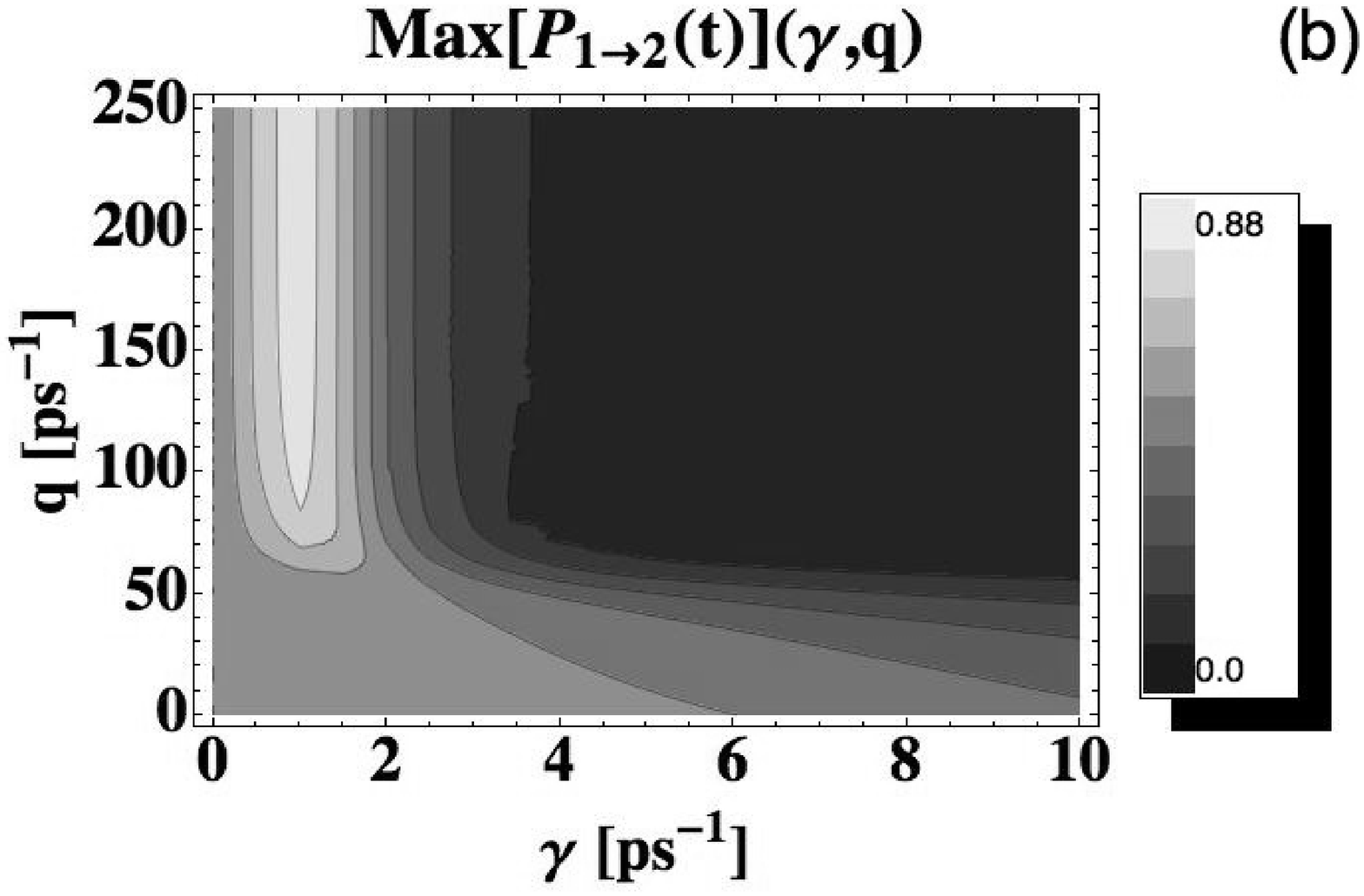}
\caption{Maximum of the probability of transition $\mathrm{Max}\left[P_{1\rightarrow2}(t)\right]$ for a dimer coupled to two spin baths as a function of
the correlations $q$ between the baths and the coupling constants $\gamma_1=\gamma_2=\gamma$ of the interaction with
the baths in the non-zero temperature case.
In Fig. 3(a) the temperature of the baths is 77K, while in Fig. 3(b) it is 300K. For both figures the parameters are $N_1=22$, $N_2=20$,
$\alpha_1=\alpha_2=250\,\mathrm{ps}^{-1}$, $J=10\,\mathrm{ps}^{-1}$ and
$\varepsilon_2-\varepsilon_1=20\,\mathrm{ps}^{-1}$.}
\end{figure}

Figure 3 addresses the question of the influence of
correlations between the environments on the energy transfer in the non-zero temperature case. It shows
the maximum of the probability of transition as a function of the coupling strengths and correlation parameter for two different bath temperatures: 77K in Fig. 3(a) and 300K in Fig. 3(b). The introduction of correlations between the baths results in an increase in the probability of transition in a well defined region of the coupling parameter. Interestingly, in the case of higher temperatures, a stronger correlation between the baths is required to observe an increase of the probability of transition. For both temperatures the maximum of the transition probability was found on timescales of the order of few hundreds of femtoseconds. It is important to stress that for the simple model considered here the maximum of the transition probability reaches 99\% at 77K and 88\% at 300K.

In conclusion, in this letter we have investigated the possible relevance of a spin environment in assisting energy transfer in a dimer system. Even for a very simple model with biologically applicable parameter ranges and timescales, it is found that the transition probability in the dimer is dramatically increased. In particular, we have demonstrated that for this class of models correlations between the environments contribute to the increase of the quantum efficiency of transport. Interestingly, for a dimer coupled to two different baths, introducing correlations between the baths improves quantum transport. These promising results motivate further study of energy transfer networks and more complex environmental models.

This work is based upon research supported by the South African
Research Chair Initiative of the Department of Science and
Technology and National Research Foundation.


\begin{thebibliography}{99}
\bibitem{toqs} H.-P. Breuer and F. Petruccione, \textit{The Theory
of Open Quantum Systems} (Oxford University Press, 2002).

\bibitem{Engel} G.S. Engel \textit{et al.}, Nature \textbf{446}, 782 (2007). 

\bibitem{pcn} G. Panitchayangkoon \textit{et al.}, Proc. Natl. Acad. Sci. U.S.A. \textbf{107}, 12766 (2010)

\bibitem{Fl} H. Lee, Y.-C. Cheng, and G.R. Fleming, Science \textbf{316}, 1462 (2007).

\bibitem{Collini} E. Collini \textit{et al.}, Nature \textbf{463}, 08811 (2010).


\bibitem{ritz}  T. Ritz, A. Damjanovic, and K. Schulten, Chem. Phys. Chem. \textbf{3}, 243 (2002).

\bibitem{Gr} R. van Grondelle and V.I. Novoderezhkin, Phys. Chem. Chem. Phys. \textbf{8}, 793 (2006).

\bibitem{wave} J.S. Briggs and A. Eisfeld, Phys. Rev. E \textbf{83}, 051911 (2011).

\bibitem{AG}   M. Mohseni \textit{et al.}, J. Chem. Phys. \textbf{129}, 174106 (2008); M.B. Plenio and S.F. Huelga, New J. Phys. \textbf{10}, 113019 (2008); P. Rebentrost \textit{et al.}, New J. Phys. \textbf{11}, 033003 (2009).

\bibitem{osc} V. Vedral and T. Farrow, arXiv:1006.3775 (2010); F.L. Semi\~ao, K. Furuya, and G.J. Milburn, New J. Phys. \textbf{12}, 083033 (2010); S. Lloyd and M. Mohseni, New J. Phys. \textbf{12}, 075020 (2010); A. Asadian \textit{et al.}, New J. Phys. \textbf{12}, 075019 (2010).

\bibitem{ent} M. Sarovar \textit{et al.}, Nature Physics \textbf{6}, 462 (2010); F. Caruso \textit{et al.}, Phys. Rev. A\textbf{81}, 062346 (2010); T. Scholak \textit{et al.}, Phys. Rev. E \textbf{83}, 021912 (2011).

\bibitem{comp} T. Renger and R.A. Marcus, J. Chem. Phys. \textbf{116}, 9997 (2002).

\bibitem{gilm} J. Gilmore and R.H. McKenzie, J. Phys.: Condens. Matter \textbf{17}, 1735 (2005).

\bibitem{ishi} A. Ishizaki and G. R. Fleming, J. Chem. Phys. \textbf{130}, 234110, (2009); A. Ishizaki and G. R. Fleming, J. Chem. Phys. \textbf{130}, 234111, (2009).



\bibitem{nonM} M .Thorwart \textit{et al.}, Chem. Phys. Lett. \textbf{478}, 234 (2009); V. I. Novoderezhkin and R. van Grondelle, Phys. Chem. Chem. Phys. \textbf{12}, 7352 (2010).

\bibitem{BBP} H.-P. Breuer, D. Burgarth, F. Petruccione, Phys. Rev. B\textbf{70}, 045323 (2004).

\bibitem{spin_spin} R. Haberkorn and M. E. Michel-Beyerle, Biophys J. \textbf{26}, 489 (1979); A. J. Hoff, Quarterly Reviews of Biophysics \textbf{14}, 4, 599 (1981); S. G. Boxer, Biochimica et Biophysica Acta \textbf{726}, 265 (1983); A. J. Hoff, Photochemistry and Photobiology \textbf{43}, 6, 727 (1986); A. J. Hoff and J. Deisenhofer, Phys. Rep. \textbf{287}, 1 (1997); E. Daviso \textit{et al}, J. Phys. Chem. C \textbf{113}, 10269 (2009); P. L. Hasjim \textit{et al}, J. Phys. Chem. B \textbf{114}, 14194 (2010).

\bibitem{molmer} J. Wesenberg and K. Molmer, Phys. Rev. A\textbf{65}, 062304 (2002).

\bibitem{yamen} Y. Hamdouni, M. Fannes and F. Petruccione, Phys. Rev. B\textbf{73}, 245323 (2006).

\bibitem{photo} J. Adolphs and T. Renger, Biophys. J. \textbf{91}, 2778 (2006)

\end{thebibliography}
\end{document}